# Boron-10 lined RPCs for sub-millimeter resolution thermal neutron detectors: Feasibility study in a thermal neutron beam

**L. M. S. Margato** [a, 1], **A. Morozov** [a], **A. Blanco** [a], **P. Fonte** [a, b], **F.A.F. Fraga** [c], **B. Guerard** [d], **R. Hall-Wilton** [e, f], **C. Höglund** [e, g], **A. Mangiarotti** [h], **L. Robinson** [e], **S. Schmidt** [e,i], **K. Zeitelhack** [j]

[a] *LIP-Coimbra, Departamento de Física, Universidade de Coimbra*
*Rua Larga, 3004-516 Coimbra, Portugal*

[b] *Coimbra Polytechnic - ISEC*
*Coimbra, Portugal*

[c] *Departamento de Física, Universidade de Coimbra*
*Rua Larga, 3004-516 Coimbra, Portugal*

[d] *ILL-Institut Laue-Langevin*
*71 avenue des Martyrs - CS 20156 - FR-38042 GRENOBLE CEDEX 9, France*

[e] *European Spallation Source ERIC (ESS)*
*P.O Box 176, SE-221 00 Lund, Sweden*

[f] *Mid-Sweden University*
*SE-85170 Sundsvall, Sweden*

[g] *Linköping University, IFM, Thin Film Physics Division*
*SE-581 83, Linköping, Sweden*

[h] *IFUSP, Institute of Physics, University of São Paulo, Brazil*

[i] *IHI Ionbond AG- Industriestraße 211*
*4600 Olten, Switzerland*

[j] *Heinz Maier-Leibnitz Zentrum (MLZ), FRM-II, Technische Universität München*
*D-85748 Garching, Germany*

ABSTRACT: The results of an experimental feasibility study of a position sensitive thermal neutron detector based on a resistive plate chamber (RPC) are presented. The detector prototype features a thin-gap (0.35 mm) hybrid RPC with an aluminium cathode lined with a 2 μm thick $^{10}B_4C$ neutron converter layer enriched in $^{10}B$ and a float glass anode. A detection efficiency of ≈ 6.2% was measured for the neutron beam ($\lambda$ = 2.5 Å) at normal incidence. A spatial resolution better than 0.5 mm FWHM was demonstrated.

---

[1] E-mail: margato@coimbra.lip.pt

## 1. Introduction

In the previous paper [1] we have introduced a new type of position sensitive neutron detector (PSND) based on thin-gap resistive plate chambers (RPC) and $^{10}B_4C$ neutron converters [2-4]. Analysis of the expected characteristics of the detector has shown potential to reach sub-millimeter spatial resolution (down to 100 μm), high detection efficiency for thermal neutrons (>50% in multilayer or inclined detector architecture) and fast timing (sub-ns temporal resolution [5]). The RPCs [6, 7] offer several practical features which are attractive for neutron detectors, such as insensitivity to magnetic fields, intrinsic discharge-protection mechanism, high modularity and scalability, and low cost per unit area. A unique combination of these features makes this type of detectors a strong candidate for applications in neutron scattering science (NSS), homeland security and geology.

This paper reports the first results of the experimental feasibility study of a PSND prototype with a thin-gap hybrid RPC. The aluminium cathode of the RPC is lined with a 2 μm layer of $^{10}B_4C$. Deposition of $B_4C$ converters, enriched in $^{10}B$, was performed at the Linköping University and in the ESS Detector Coatings Workshop in Linköping. The prototype, constructed at LIP-Coimbra lab was installed at the CT2 monochromatic thermal neutron beam (λ = 2.5 Å) at the Institut Laue-Langevin (ILL).

The detector tests have demonstrated sensitivity to thermal neutrons with a wide plateau. The detection efficiency of the prototype with a single neutron converter layer was measured and compared with the results of Monte Carlo simulations. Scans of the detector with a narrow neutron beam were performed and sub-millimeter spatial resolution of the detector was demonstrated.

## 2. Experimental setup

### 2.1. Hybrid RPC with $^{10}B_4C$ neutron converter

The feasibility of using $^{10}B$-lined RPCs for PSNDs was evaluated in this study with a single-gap RPC prototype in hybrid configuration [1]. Figure 1 shows a schematic drawing of the detector. The sensitivity to thermal neutrons originates from the neutron capture reaction $^{10}B(n,^4He)^7Li$. The $^7Li$ and $^4He$ fission fragments exiting the converter into the gas-gap with sufficient energy generate ionization followed by Townsend avalanches. A detailed description of the working principle of this detector can be found in the previous paper [1].

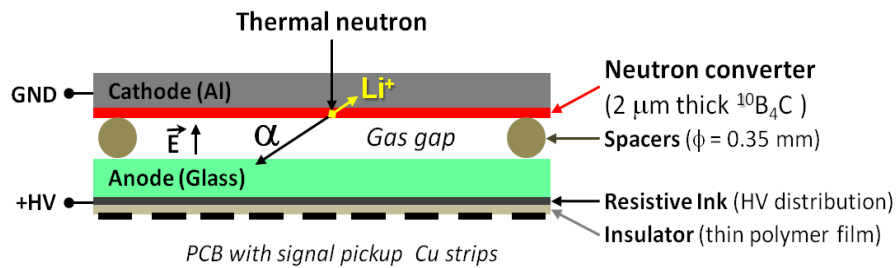

**Figure 1.** Schematic drawing of a single-gap hybrid RPC lined with a $^{10}B_4C$ neutron converter.

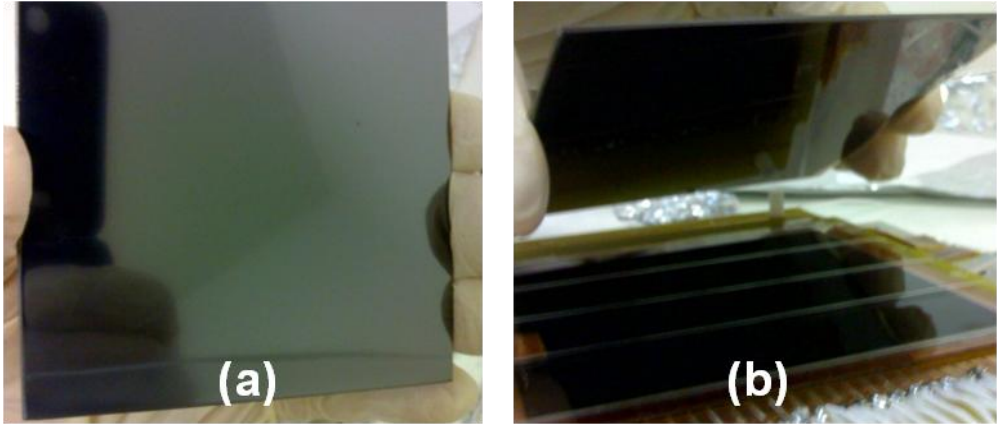

**Figure 2.** Photographs of the aluminium cathode plate lined with 2 μm thick layer of $^{10}B_4C$ (a), and of the cathode lifted above the anode glass plate (b). The RPC electrodes are separated by 0.35 mm diameter nylon monofilaments (visible as the white lines on top of the anode) defining a uniform gas-gap width.

The aluminium cathode (1 mm thick, $8 \times 8$ cm$^2$ area) was lined with a 2 μm thick layer of $^{10}B_4C$ on the surface facing the gas-gap. The $^{10}B_4C$ coating with $^{10}B$ enrichment of >97% was deposited onto one side of the aluminium plates at the Linköping University and in the ESS Detector Coatings Workshop in Linköping. The deposition was performed at an industrial-scale DC magnetron sputtering system at a substrate temperature bellow 100ºC to avoid deformation of the aluminium due to residual stresses in the coating. Further details on the deposition technique and parameters can be found in [2-4].

The float glass anode plate (0.35 mm thick, $8 \times 8$ cm$^2$) was lined on the surface opposite to the gas-gap with a thin layer of a resistive ink. The function of the ink is to evenly distribute the high voltage (HV) over the entire electrode surface. Nylon monofilaments (0.35 mm diameter) were placed between the anode and the cathode plates (see figure 2) to define the gas-gap width.

The array of signal pick-up electrodes (2 mm wide copper strips with a pitch of 2.5 mm) was engraved on a 1.5 mm thick printed circuit board (PCB). The PCB was installed with electrodes facing the anode resistive ink, insulated from it with a 50 μm thick acetate foil (see figure 1).

### 2.2. Detector layout and setup at a neutron beamline

Two hybrid RPCs were assembled, identical in all aspects except that the first one (RPC-1) did not have a neutron converter, while the second one (RPC-2) had a 2 μm layer of $^{10}B_4C$ deposited on the cathode. Both RPCs were installed inside the same gas cell as shown in figure 3.

The cell was equipped with two 3 mm thick aluminium windows on opposite sides, each facing the cathode of the corresponding RPC. This arrangement allowed to operate both RPCs in the identical conditions with respect to the neutron beam irradiation by rotating the detector 180 degrees around the vertical axis.

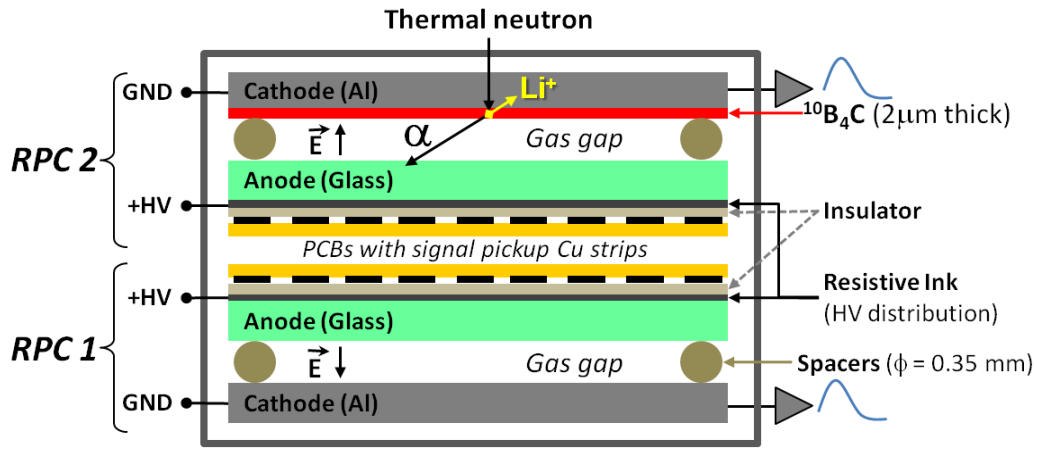

**Figure 3.** Schematics drawing of the detector gas cell enclosing two single-gap hybrid RPCs: RPC-1 does not have a neutron converter, while the cathode of the RPC-2 is lined with a 2 μm thick layer of $^{10}B_4C$.

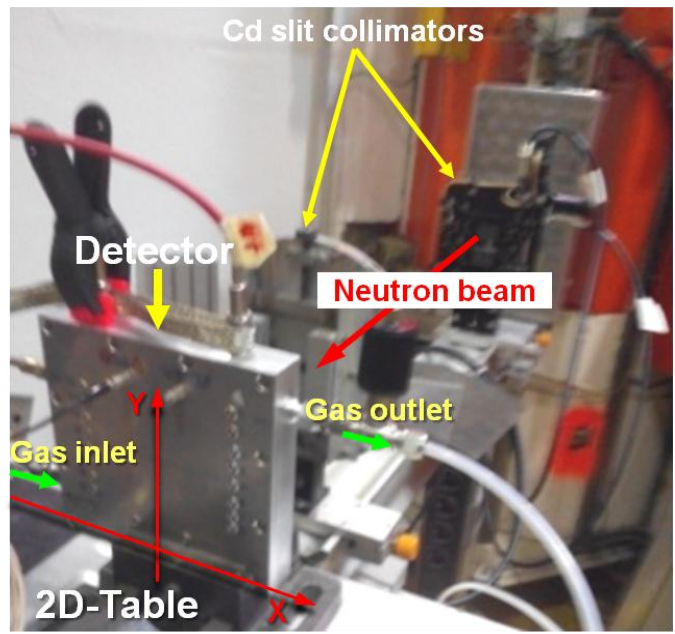

**Figure 4.** Detector installed on a 2D moving stage on the CT2 neutron beam line (λ = 2.5 Å) at ILL. The beam was collimated by a pair of orthogonal cadmium slits (vertical and horizontal directions) mounted on an optical bench.

The gas cell was filled with tetrafluoroethane ($C_2H_2F_4$) at the atmospheric pressure. The gas had industrial purity grade and was supplied by Linde. All experiments were performed in this study with the gas flow rate of ≈ 2 cc/min and ambient temperature of 25ºC.

The gas inlet and outlet valves were mounted on the opposite corners of the cell (see figure 4) in order to ensure an efficient circulation of the gas through the entire volume. All parts of the gas system were interconnected with polytetrafluoroethylene (PTFE) tubes.

Polarization potential up to 2.8 kV was applied to the anodes using a CAEN N471A power supply. The total current was monitored using the integrated gauge (1 nA resolution).

The setup was installed on the CT2 monochromatic thermal neutron beamline (λ = 2.5 Å) at ILL (see figure 4). The gas cell was positioned on a 2D moving stage. Two mutually-perpendicular cadmium slits of adjustable width were used to define the irradiated area. The cell was always oriented with the beam incident orthogonally to the entrance window.

After installation at the beam line, the gas cell was flushed with the working gas at a flow rate of about 10 cc/min for 4 hours. Before starting measurements, the RPCs were kept polarized at 2.8 kV for 12 hours. The dark current measured after this procedure was below 1 nA.

### 2.3. Electronic readout and DAQ system

For the neutron sensitivity measurements the signals induced in the RPC cathode plate, kept at the ground potential, were readout by a charge sensitive preamplifier (sensitivity of 1 V/pC) and then fed into a linear amplifier (Canberra 2021) with the shaping time of 1 µs. The output of the amplifier was connected to a single-channel analyzer (SCA, ORTEC model 553). Finally, the digital output of the SCA was read by a counter module. The lower level threshold of the SCA was adjusted to suppress the electronic noise. The calibration of the entire signal processing chain was performed using an ORTEC 448 pulser and a 2 pF calibrated capacitor to inject a known charge in the input of the preamplifier.

The signal pick-up strips, providing the neutron event position information, were readout with a custom data acquisition system based on the MAROC chip [8]. MAROC3 features 64 inputs, each equipped with a low impedance preamplifier with adjustable gain and a slow shaper with configurable time constant. The charge was digitized by the on-chip Wilkinson ADC (12 bits). The slow shaper was set to the maximum possible integration time (time constant of ~150 ns), resulting in collection of ≈15% of the total charge induced in the RPC pick-up electrodes during a neutron event (see figure 5).

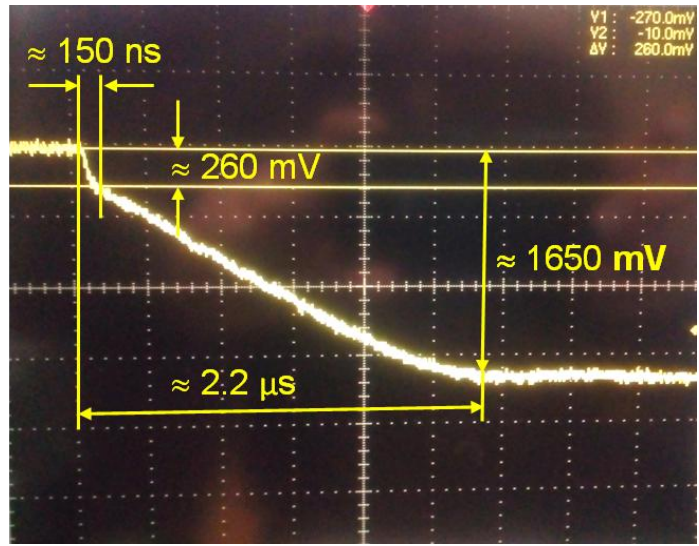

Figure 5. Cathode pulse at the output of an inverting charge preamplifier shown at the screen of a Tektronix TDS7104 oscilloscope (amplitude and time scales of 500 mV/div and 400 ns/div, respectively). At the beginning of the pulse there is a fast electronic component followed by the slow ionic one. Integration during the first 150 ns results in collection of ≈ 15% of the total induced charge.

For practical reasons only 16 signal pick-up strips in the central area of the PCB were connected to the acquisition system. The remaining strips, 8 on the left side and 8 on the right, were grounded. The digitized signals were transferred to a PC via a USB connection and stored for further processing. After removing pedestals, the signals from three adjacent strips showing the strongest signals were used to calculate the position of the neutron capture by applying the center of gravity (CoG) algorithm [9].

## 3. Results and discussion

### 3.1. Plateau measurement

The sensitivity to thermal neutrons as a function of the polarization voltage was measured for both RPCs (RPC-1 without and RPC-2 with the neutron converter, see section 2.2) using a monochromatic ($\lambda$ = 2.5 Å) neutron beam.

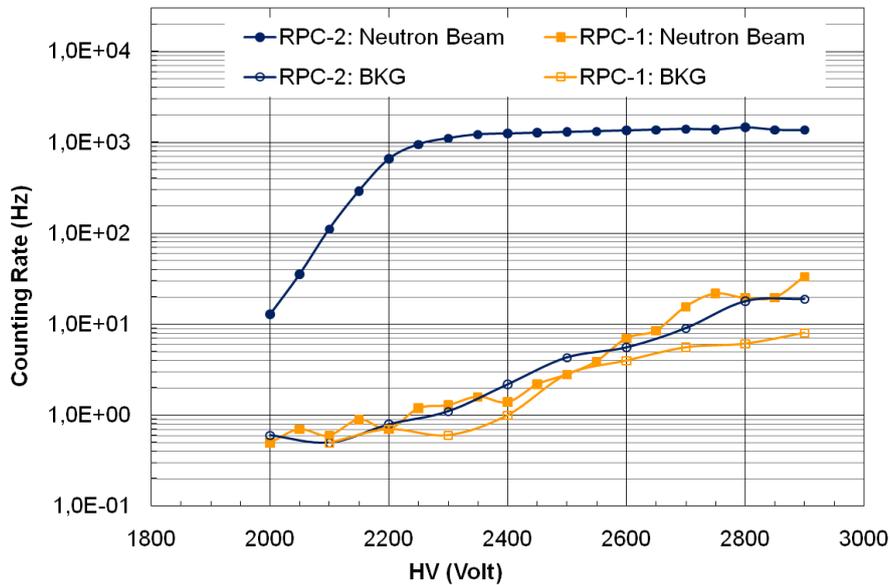

**Figure 6.** Counting rate recorded with the RPC-1 and RPC-2 as a function of the polarization voltage with and without the neutron beam irradiation. Square and circular markers designate the results for the RPC-1 and RPC-2, respectively. In both cases the markers are filled when the beam is on, and are empty when the beam is off.

Figure 6 shows the counting rate as a function of the polarization voltage with and without the neutron beam irradiation. The beam intensity was attenuated by 9 mm of acrylic glass to avoid saturation (see section 3.2). The irradiated area on the detector was $2 \times 4$ cm$^2$, which corresponds to the largest possible neutron beam cross-section area available with this setup. Since the signals have been read on the metallic cathode, the background count rate has the contribution of the entire active area of the RPC ($8 \times 8$ cm$^2$).

The counting rate of the RPC-2 recorded with the beam exhibits a strong increase with the polarization potential until $\approx$ 2.3 kV and then stabilizes showing a wide plateau well above the counting rate recorded without the beam. On the other hand, the counting rates of the RPC-1 recorded with and without the beam are very similar, confirming insensitivity of the RPC without the neutron converter to thermal neutrons. This difference in the response to the beam demonstrates the feasibility of using hybrid $^{10}$B-RPCs for thermal neutron detection.

The fact that the counting rate of the RPC-1 in the presence and absence of the beam is essentially the same (considering the uncertainties) for polarization voltages up to 2.5 kV (see figure 6) also suggests that the RPC has a low sensitivity to the gamma rays produced by the nuclear interactions of neutrons with the materials of the detector and experimental setup, most importantly the cadmium slits. For potentials above 2.5 kV the counting rate recorded with the beam is higher than the one without the beam, and the difference increases with the potential, which can be explained by an increase in the sensitivity of the RPC to the gamma rays generated due to the neutron beam irradiation. Note that the neutron sensitivity plateau is reached with the RPC-2 at a significantly lower potential ($\approx$ 2.3 kV).

A background counting rate of less than $2\times10^{-2}$ Hz/cm$^2$ were measured for both RPCs operated with a voltage of 2.4 kV. For the voltages above 2.5 kV the difference in the background level of the RPC-1 and RPC-2 (figure 5) is most likely explained by imperfections of the RPC electrode surfaces [10, 11].

The RPC counts recorded without the beam originates from two main processes. The first one is spontaneous generation of avalanches in the gas-gap appearing, for example, in the vicinity of micro-defects at the electrodes surface. The second one is energy deposition due to the radiation background, which includes cosmic rays, fast neutrons leaking from the reactor and fission fragments from the decay of radioactive impurities present in the materials of the RPC electrodes. The background counting rate of the RPCs can be reduced using several approaches. One is to improve the surface quality of the glass and aluminium plates [12]. Another one is to optimize the working gas mixture minimizing the number of streamers [13-15]. Also, several strategies for reducing the background connected to the presence of traces of the radioactive isotopes in aluminium alloys [16] for $^{10}$B based thermal neutron detectors were reported in [17]. These strategies include usage of radiopure materials or, alternatively, introduction of a thin layer of nickel at the aluminium surface in order to stop the alpha particles produced in the decay chain of the radioactive impurities, such as, e.g., traces of thorium and uranium.

### 3.2. Detection efficiency

The RPC-2 counting rate as a function of the neutron beam flux was recorded to establish the range where the detector is operating in linear mode. The flux was adjusted by changing the number of acrylic glass attenuators placed in front of the first cadmium slit (see section 2.2). The RPC was operated at 2.4 kV, which is well within the plateau (see figure 6).

The results are shown in figure 7 for two areas (1 and 8 cm$^2$) irradiated with the beam. For comparison, the figure also gives the counting rate recorded with a $^3$He-tube (90% efficiency at 2.5 Å) for 1 cm$^2$ beam area. The choice of 1cm$^2$ area was made to avoid loss of the tube detection efficiency due to its geometry (diameter is 5 cm), while 8 cm$^2$ is the maximum beam area available at the beam line. Note that the counting rate of the RPC-2 scales linearly with the irradiated area (factor of $\approx$ 8).

Figure 7 demonstrates that the counting rate measured with the RPC-2 remains linear up to $\approx$ 1 kHz/cm$^2$. Above this value the RPC-2 starts losing efficiency, which is expected taking into account that the resistive anode was made from common float glass, that have quite high

– 7 –

resistivity (~ $10^{12}$ Ω·cm). Similar counting rates (1 - 3 kHz/cm$^2$) were already reported for float glass RPCs designed to detect minimum ionizing particles (MIPs) [18, 19].

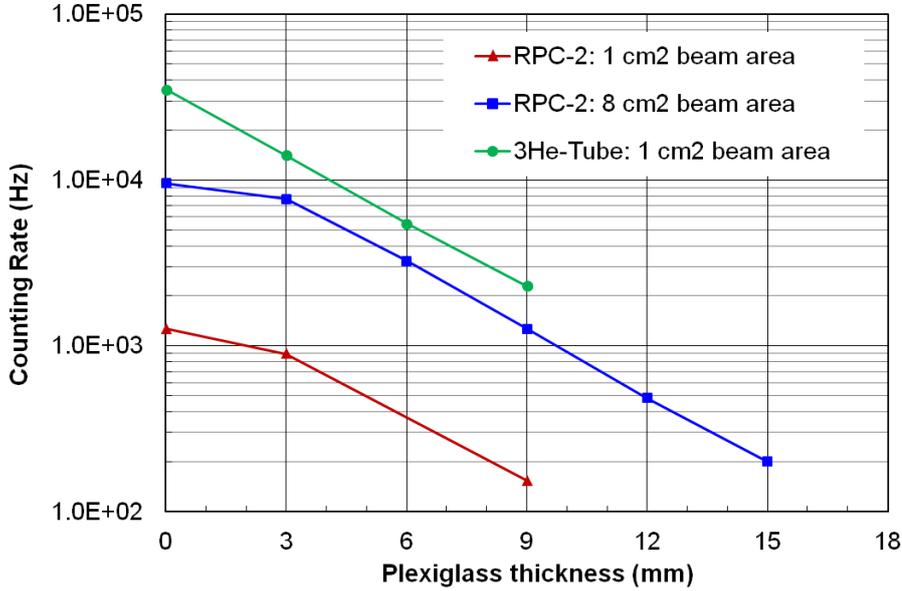

**Figure 7.** Counting rate, measured with the RPC-2 and the $^3$He tube, as a function of the combined thickness of the acrylic glass attenuators. The RPC-2 data are shown for two irradiated areas (1 and 8 cm$^2$); the data for the $^3$He tube were recorded with the area of 1 cm$^2$.

The detection efficiency was calculated as the ratio of the number of detected neutron events per second, corrected for the background, to the neutron beam flux incident on the detector's sensitive area. The flux was measured with the $^3$He tube. The background counting rate was recorded in the same conditions but in the absence of the beam. For the irradiated area of 1 cm$^2$ and the combined attenuator thickness of 9 mm, we obtain the neutron detection efficiency of the RPC-2 of 6.17 % ± 0.64%.

This efficiency value is quite close to the value of 6.1% obtained for 2.5 Å neutrons in simulations of the detector with the ANTS2 toolkit [20] (see [1] for the description of the simulation procedure).

As expected, the detection efficiency of the RPC with a single converter layer orientated orthogonally to the neutron beam is quite low. However, as described in [1], PSNDs based on the $^{10}$B hybrid RPC with significantly higher efficiencies can be designed using a multilayer configuration or orienting the RPCs in a small angle with respect to the neutron beam.

### 3.3. Spatial resolution

For evaluation of the spatial resolution of the detector prototype, the collimators were adjusted to define a narrow beam of 0.5 mm by 40 mm (X and Y directions, respectively). The RPC polarization potential was 2.6 kV (cathode grounded) and the neutron beam was attenuated by 6 mm of acrylic glass, assuring operation of the detector in liner regime (see section 2.2).

The X positions of the neutron capture events were calculated with the centroid algorithm taking into account the signals from the three neighboring strips with the strongest signals. The reconstruction results for three positions of the detector (step of 0.5 mm in X direction) are presented in figure 8.

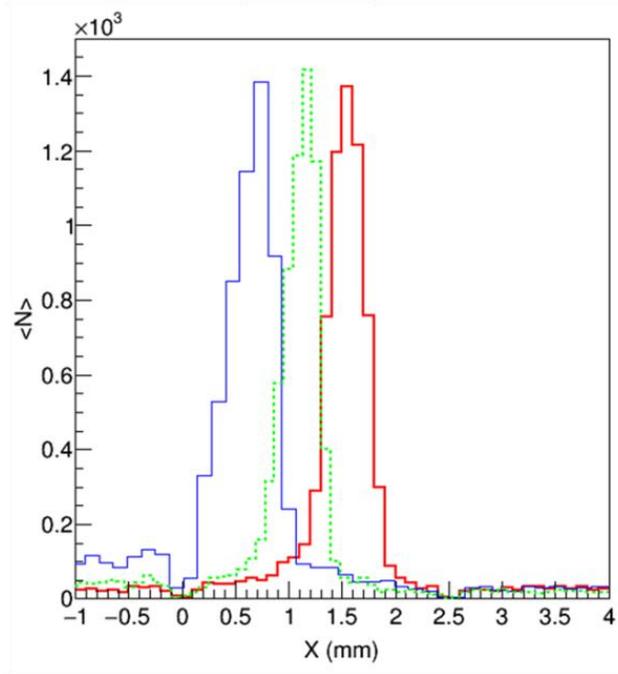

**Figure 8.** Distribution of the reconstructed X positions of the neutron events recorded with the RPC-2 at three different locations (0.5 mm step in X direction). The neutron beam width is 0.5 mm and the pick-up electrode pitch is 2.5 mm.

The distributions of the reconstructed positions shown in figure 8 suggest that the spatial resolution is ≤ 0.5 mm (FWHM). Note that the intrinsic resolution of the detector should be significantly better taking into account that the beam width is 0.5 mm.

The profile of the distribution of the reconstructed positions (figure 8) is different from the shape one might expect: a uniform distribution (0.5 mm FWHM) convoluted with a narrow bell-shaped curve. The observed distribution appears mainly due to the distortions introduced by the centroid reconstruction algorithm, which assumes linear dependence of the induced signal in a pick-up electrode with the distance to the event position. In contrast to this assumption, the strip response function (SRF), describing this dependence, can have a complex profile and strongly depends on the spread of the signals in the layer of resistive ink [21] as well as on the RPC geometry. A significant improvement in the reconstruction fidelity can be achieved by applying a correction function to the results of the centroid reconstruction [22] or by using more sophisticated reconstruction methods, such as, e.g., statistical reconstruction [23]. Both approaches require calibration data recorded by scanning the detector with a narrow neutron beam using a small step. Another factor which can affect the distribution profile is neutron scattering on the collimator, entrance window and inside the detector.

Note that the offset of the centroid of the induced charge from the true neutron capture position is expected to be very small for this detector type due to the fact that the gas-gap is very narrow (0.35 mm) and the anti-parallax effect of the electron avalanche development in RPCs described in [1].

The results presented here suggests that by optimizing the RPC parameters, the geometry of the signal pick-up electrodes, the front end electronics and the reconstruction algorithm, it seems to be possible to achieve a spatial resolution better than 100 μm as it was already demonstrated for RPC detectors for MIPs [24].

## 4. Conclusions

This study demonstrates the feasibility of using thin-gap RPCs with $^{10}B_4C$ neutron converters for position sensitive thermal neutrons detectors. For the detector prototype with a single converter layer the detection efficiency measured with a monochromatic neutron beam (λ= 2.5 Å) at normal incidence is ≈ 6.2%, which agrees well with the results (≈ 6.1 %) of the Monte Carlo simulations [1].

The detector was irradiated by a narrow (0.5 mm) neutron beam and the positions of the detected events were reconstructed with the CoG method. The results demonstrate that the spatial resolution of the detector is better than 0.5 mm FWHM, confirming the predictions made in [1].

The fast timing characteristic of RPCs [25] should allow to measure the neutron time-of-flight (TOF) with nanosecond time resolution, which makes this type of detector also well suited for energy-selective or fast dynamics neutron imaging.

The wide plateau in the counting rate as a function of the polarization voltage appears at the potentials significantly lower than the ones used in RPC-based MIPs detectors. This fact suggests that the gamma sensitivity of the detector should be low, but the exact value has to be established in a dedicated study.

The anode of the prototype tested in this study was build from a common float glass with high resistivity (~ $10^{12}$ Ω·cm), which strongly limited the maximum counting rate of the detector (~1 kHz/cm$^2$). Using lower resistivity materials and lower operational voltages should increase this value by one or two orders of magnitude [25-27].

As discussed in [1], RPCs based PSNDs in multilayer and inclined geometries should be able to operate at significantly high counting rates (e.g. the counting rate scales nearly linearly with the number of layers or with inverse sine of the angle of incidence) and provide detection efficiency exceeding 50%. Currently we are working on characterization of the next generation prototype featuring $^{10}B_4C$ lined double-gap RPCs in a multilayer architecture.

## Acknowledgments

This work was supported in part by the European Union's Horizon 2020 research and innovation programme under grant agreement No 654000. Richard Hall-Wilton, Carina Höglund, Linda Robinson, and Susann Schmidt would like to acknowledge the financial support of the EU H2020 Brightness Project, grant agreement 676548.